\documentclass[12pt]{article}

\usepackage{amsmath}
\usepackage{amssymb}

\usepackage{graphicx}

\usepackage{cite}

\usepackage{color} 

\usepackage[labelfont=bf,labelsep=period,justification=raggedright]{caption}

\makeatletter
\renewcommand{\@biblabel}[1]{\quad#1.}
\makeatother

\date{}

\begin{document}

\begin{flushleft}
{\Large
\textbf{Binary regression analysis with network structure of respondent-driven sampling data}
}
\\ \vspace{6pt}
Leonardo S. Bastos$^{1,\ast}$, 
Adriana A. Pinho$^{2}$, 
Claudia Code\c{c}o$^{3}$, 
Francisco I. Bastos$^{4}$
\\ \vspace{6pt}
$^1$ Departamento de Estat\'{\i}stica, Universidade Federal Fluminense, Niter\'oi, RJ, Brazil \\
$^2$ Escola Nacional de Sa\'ude P\'ublica, Funda\c{c}\~ao Oswaldo Cruz, Rio de Janeiro, RJ, Brazil \\
$^3$ Programa de Computa\c{c}\~ao Cient\'{\i}fica, Funda\c{c}\~ao Oswaldo Cruz, Rio de Janeiro, RJ, Brazil \\
$^4$ Instituto de Comunica\c{c}\~ao e Informa\c{c}\~ao Cient\'{\i}fica e Tecnol\'ogica em Sa\'ude, Funda\c{c}\~ao Oswaldo Cruz, Rio de Janeiro, RJ, Brazil \\
$^\ast$ Corresponding author. Email: lbastos@est.uff.br
\end{flushleft}

\begin{abstract}
Respondent-driven sampling (RDS) is a procedure to sample from hard-to-reach populations. It has been widely used in several countries, especially in the monitoring of HIV/AIDS and other sexually transmitted infections. Hard-to-reach populations have had a key role in the dynamics of such epidemics and must inform evidence-based initiatives aiming to curb their spread. In this paper, we present a simple test for network dependence for a binary response variable. We estimate the prevalence of the response variable. We also propose a binary regression model taking into account the RDS structure which is included in the model through a latent random effect with a correlation structure. The proposed model is illustrated in a RDS study for HIV and Syphilis in men who have sex with men implemented in Campinas (Brazil).
\end{abstract}

{\bf Keywords:} Respondent-driven sampling, Binary regression, Network dependence.

\section{Introduction}

Respondent-driven Sampling, RDS, was originally formulated by Heckathorn \cite{Hec97} as comprising a first-order Markovian process, supposed to reach equilibrium after a given number of waves (originally estimated as six). It is a sampling scheme used to access hard-to-reach populations, e.g. heavy drug users \cite{Saletal11a}. RDS has been widely used in several countries and well-known public health institutes \cite{Maletal08}.

More recent developments understand RDS as a Markov Chain Monte Carlo, formally defined by \cite{GoeSal09} and later used by the authors in a comprehensive series of simulations \cite{GoeSal10}.

An alternative perspective, to be fully developed yet, understands the networks obtained by RDS as a branching process that violates the basic assumptions of Markovian processes \cite{Pooetal09} and propose the use of stochastic context-free grammars to analyze databases generated by RDS.

Whatever the alternative to be taken in the analysis of RDS-based data, there is nowadays a consensus that RDS constitutes a powerful strategy to assess hard-to-reach population, such as crack users, for whom RDS generates samples which are substantially different from those based on institutional random samples \cite{Oteetal12}. In the same way, some biases traditionally associated with chain-referral samples, such as those secondary to the role of the so-called ``super-recruiters'', whose very existence it is blocked by the establishment of a priori recruiting quotas in the context of RDS studies \cite{Tif06}.

In order to learn about the prevalence of specific characteristic of the population different estimators have been developed for RDS \cite{Hec97, Hec02, SalHec04, VolHec08, Gil09}. However, there is little research on estimating risk factors for hard-to-reach population taking into account the RDS approach. In this paper, we propose a strategy for carrying out regression analysis for RDS data.

\section{Methods}

\subsection{Respondent-driven sampling}

Heckathorn \cite{Hec97} proposed the use of a snowball sampling method \cite{Goo61} to sample from hard-to-reach populations. The proposed sampling scheme is called Respondent driven sampling, or simply RDS. In a snowball sampling, the data are collected according to a chain-link recruitment process where few participants, called seeds, are chosen from the population of study, these participants are asked to recruit future participants of the same population group, which will be asked to recruit future participants of the same population, and so on. This process forms a network of recruits. In a respondent-driven sampling, the participants also provide information about their personal network size and each individual has a unique number or code allowing us to connect recruiters and participants.



In general, we are interested in some quantities (or variables) associated with each participant of the sample. These quantities may be influenced by the interaction among the participants. We call this association as network dependence. In this sense, the variance within a recruiter-recruitee dyad (pair) tend to be different (more likely to be less pronounced) than the variance between two interviewees not connected by a given referral link. \cite{McPheetal01}.


If a quantity of interest is a categorical variable, then it is possible to built a contingency table with the recruiter values in the columns and the participant values in the rows. If there is some dependence between the status of the quantity for the recruiters and the status of the quantity for the participant, then it may suggest a network dependence. Therefore, a Pearson $\chi^2$ independence test for contingency tables can be used for checking evidence of network dependence.


\subsubsection*{RDS estimators}

Suppose we are interested in a characteristic $A$ which can be observed or not in each individuals. The simplest estimator of the prevalence of the characteristic $A$, $\theta_A$, is the naive estimator given by
\begin{equation} \label{naive}
\hat{\theta}_A^{(n)} = \frac{n_A}{n}
\end{equation}
where $n$ is the sample size, $n_A = \sum_{i = 1}^{n} \hbox{1}{\hskip -2.5 pt}\hbox{I}_A(i)$ and $\hbox{1}{\hskip -2.5 pt}\hbox{I}_A(i)$ is an indicator function where $\hbox{1}{\hskip -2.5 pt}\hbox{I}_A(i)=1$ if the individual $i$ presents the characteristic $A$ and $\hbox{1}{\hskip -2.5 pt}\hbox{I}_A(i)=0$ otherwise. If the sample size is big and independence among individuals is a reasonable assumption, then the 95\% confidence interval is $(\hat{\theta}_A^{(n)} \pm 1.96 (\hat{\theta}_A^{(n)} (1-\hat{\theta}_A^{(n)})/n)^{1/2})$. However, the independence assumption may be a strong assumption for respondent-driven sampling data. 

An estimator that takes into account the network structure was proposed \cite{Hec97}, it is called RDS I. The RDS I estimator was improved by  Volz and Heckathorn \cite{VolHec08}. The estimator is called RDS II, and it is given by the following
\begin{equation}\label{rds2}
\hat{\theta}_A^{(\mbox{RDS II})} = \frac{\sum_{i = 1}^{n} \hbox{1}{\hskip -2.5 pt}\hbox{I}_A(i) \delta_i^{-1}}{\sum_{i = 1}^{n} \delta_i^{-1}},
\end{equation}
where $\delta_i$ is called the degree, and it is defined by the number of `friends´ from the same population that participant $i$ declares to have. The authors provide an estimator for the variance of (\ref{rds2}). The authors also provide a simulation study showing that the confidence intervals built using the RDS II estimator are better, in terms of average coverage probability, than the confidence intervals built with the naive estimator. Recently, another estimator has been developed called RDS III, \cite{Gil09}. However, this estimator is not explored in this paper.



\subsection{Binary regression}

Let $Y_i$ be a variable representing a characteristic of interest of the $i$-th individual interviewed in a respondent-driven sample, where $Y_i=1$ if the characteristic of interest is observed on individual $i$, and $Y_i=0$ otherwise for $i=1,2,\ldots,n$.

Risk factors can be incorporated in a binary regression model as the following 
\begin{eqnarray} \label{iid} \nonumber
Y_i & \sim & Bernoulli(\theta_i), \\
g(\theta_i) & = & \eta_i = \mathbf{x}_i^T \beta,  \quad i=1,2,\ldots,n
\end{eqnarray}
where $\mathbf{x}_i$ is a vector of possible risk factors, $\beta$ are the risk effects and $g(\cdot)$ is a link function. If the link function is the logit function, $g(z) = logit(z) = \log(z/(1-z))$, then the regression is called logistic regression.

However, the model (\ref{iid}) is valid only when the characteristic of interest is independent among the individuals in a RDS study. This is only valid when there is no network dependence.

If the contact network is known, then a latent term can be included in the logistic model where the network structure will be taken into account. This is done using a latent Gaussian Markov random field, i.e. 
\begin{eqnarray} \label{car} \nonumber 
Y_i & \sim & Bernoulli(\theta_i), \\
g(\theta_i) & = & \eta_i = \mathbf{x}_i^T \beta + \omega_i,  \quad i=1,2,\ldots,n
\end{eqnarray}
where $\omega_i$ is a latent effect of the network structure. The latent effects are modeled using the following conditional auto-regressive model, CAR, proposed by \cite{Bes74},
\begin{equation} 
\omega_i | \omega_j, i\neq j, d, \tau \sim N\left( \frac{1}{d + n_i}\sum_{i\sim j} \omega_j, \frac{1}{(n_i+d)\tau} \right),
\end{equation}
$n_i$ is the number of contacts of individual $i$ (number of connections), $i \sim j$ means the set of individuals connected to $i$,  $\tau$ is a precision hyper-parameter and $d$ is a diagonal parameter. In order to complete the model we set vague priors for $\beta$, $\tau$ and $d$.

The model (\ref{car}) is a well-known model in Bayesian spatial statistics, where neighborhood regions are considered as connections \cite{BesYorMol91}. The inference is based on the marginal posterior distributions of each parameter. These posterior marginal distributions are obtained using the integrated nested Laplace approximation, INLA, \cite{Rueetal09}. Model comparison is done by using the deviance information criterion, DIC, \cite{Spietal02}.

\section{Application: HIV and Syphilis of MSM population in Campinas, Brazil}

The RDS study carried out by de Mello et al. (2008) \cite{deMeletal08} was the first large RDS study implemented in Brazil in a single location (Campinas, Sao Paulo state). This study was part of a comprehensive initiative launched by Horizons-USAID aiming to better assess the HIV/AIDS epidemic among gay men, worldwide, using new methods targeting hard-to-reach populations \cite{Geietal10}.

\begin{figure}[htb]
\begin{center}
\resizebox*{0.85\textwidth}{!}{\includegraphics{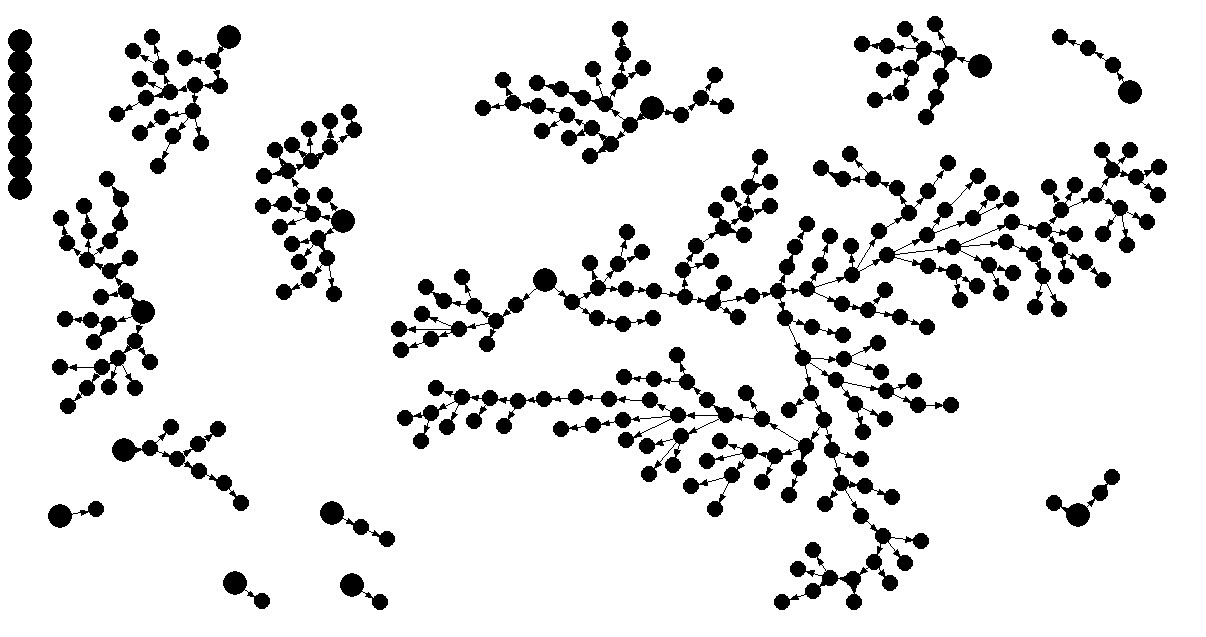}}%
\caption{Recruitment pattern of men who had sex with other men. The larger circles represent seeds and smaller circles represent subsequent recruitees.} \label{hsh.fig}
\end{center}
\end{figure}

The study comprised 658 men who have sex with men (MSM) and was preceded by a comprehensive formative study. The inclusion criteria for a participant are  (i) born male;  (ii) had anal or oral sex with another man or transvestite in the past six months; (iii) 14 years of age or older; (iv) reside in Metropolitan area of Campinas. Participants were compensated for enrolling in the study  and for each eligible man they successfully recruited into the study. The maximum number of referrals was 3. Some recruitment waves were exceedingly long, comprising over 20 successive recruitment waves. In this sense, according to the RDS original formulations, equilibrium should be reached. Figure \ref{hsh.fig} presents the observed network in the Campinas RDS study. The initial recruitment started with 10 seeds. Seven additional seeds were added 4-6 months after the study started due to slow recruitment. Another six seeds were added (eight months after the study started). Additionally, seven potential participants who arrived after the 10th month at the study site without a coupon were treated as seeds.

Subsequent, larger, Brazilian studies, were conceived as multicity studies. 
As such, they deal with a pool of local networks instead of a single, larger network. The continental size of Brazil, besides its pronounced geographic and social heterogeneity makes the analysis of such pooled data and respective weighting a formidable challenge. As shown by a former paper by our research group \cite{Toletal11}, even considering a single city (Rio de Janeiro) belonging to this multicity study (which comprised 10 cities all over the country, as of 2009-2010), we made evident structural bottlenecks (secondary to structural violence affecting Rio de Janeiro's drug scenes, \cite{Basetal07}) that hampered the very progress of the recruiting process. A posterior analysis of the geographic dimensions of another RDS-based study, carried out in Uganda's villages \cite{McCreetal11}, did not confirm our findings, much probably due to the striking social, geographic and demographic differences between Uganda's villages and the violence-laden large metropolitan scenes 
where the Rio de Janeiro's study took place.

Whatever the underlying reasons associated with these and other discrepancies, we chose here to profit from a one-site large study with gay men. 
Although homophobic crimes and other sexual identity and racial-driven crimes do unfortunately exist in Brazil (as described by the participants themselves; \cite{deMeletal08}), the gay scene in Campinas (as well in other major Brazilian urban areas) can be defined as an open scene, not affected by the same structural bottlenecks disrupting Rio de Janeiro drug scenes to the point of making some of them impervious to different attempts of researchers and health professionals to work in partnership with local leaderships and native outreach workers.

On a side note, one must observe that the first RDS (and to the best of our knowledge, so far, only one) simulation study on the accuracy of RDS I estimator was parametrized after the same data from de Mello's study \cite{Albetal11}.



\subsection{Testing the network dependence and estimating prevalences}

Table \ref{hivdep.tbl} is a contingency table of the HIV serostatus for the participants and its recruiters. 
The Pearson $\chi^2$ independence test rejects the independence null hypothesis, p-value $<$0.0001, suggesting there is evidence of network dependence for HIV serostatus.
Analogously, Table \ref{syphdep.tbl} is a contingency table of the syphilis serostatus for the participants and its recruiters. The Pearson $\chi^2$ independence test rejects the independence null hypothesis, p-value = 0.0014, suggesting there is evidence of network dependence for syphilis serostatus. Therefore, we have some evidence that we should include the network structure to estimate the prevalence and to find risk factors associated with HIV and syphilis for the MSM population in Campinas.

Table \ref{prev.tbl} provides the estimated prevalences using the naive estimator (\ref{naive}) and the RDS II estimator (\ref{rds2}), and their correspondent 95\% confidence interval. Since there is some evidence on network dependence for HIV and syphilis serostatus, the estimated prevalences that should be considered are those using the RDS estimator, i.e 7.1\% (4.7; 9.6) for HIV and 9.4\% (1.5; 17.4) for syphilis.


\subsection{Regression analysis}

In order to obtain risk factor for HIV and syphilis serostatus, we use a binary regression model with logistic link function. Hence, we have two different models: the usual logistic regression (\ref{iid}), LogReg, and the logistic regression with latent network effect (\ref{car}), NetLogReg.

The results for HIV serostatus are summarized in Table \ref{hivreg.tbl}. The DIC suggests that the logistic regression with latent network effect is better, which agrees with the fact that there is some evidence of a network dependence in this data. Although we estimated the regression coefficients, the results are interpreted as odds ratios (OR). 
Participants that received any educational material in the past 12 months are three times the odds of having HIV (OR=$3.03$, 95\%CI 1.21-8.42) compared to those who did not received any educational material. 
Participants older than 25 years are four times the odds of having HIV (OR=4.11, 95\%CI 1.68-10.62) compared to those younger than 25 years. Participants with no more than high school degree had almost three times the odds of having HIV (OR=2.91, 95\%CI 1.04-9.22) than those participants with at least college education. 
Participants who did not engage UIAI (unprotected insertive anal intercourse) in the last two months had 2.7 times the odds of having HIV (OR=2.70, 95\%CI 1.08-7.46) than those participants who engaged UIAI in the last two months.

The results for Syphilis serostatus are summarized in Table \ref{syphreg.tbl}. The DIC suggests that there is no significant difference between the two models. This seems to contradict the dependence test, however due to missingness, the sample size had to be reduced 658 to 545 participants. This reduction on the sample size removed several social observed connections leading to a weaker network dependency. Therefore, for this sample, the usual logistic regression was chosen to described the factors related to syphilis serostatus.

Participants that declared themselves transvestite had 2.49 times the odds of having syphilis (OR=2.49, 95\%CI 1.03-5.77) compared to those who declared themselves men. Participants older than 25 years had 2.93 times the odds of having syphilis (OR=2.93, 95\%CI 1.54-5.73) compare to those younger than 25 years. Participants that live in Campinas city had 2.48 times the odds of having syphilis (OR=2.48, 95\%CI 1.02-6.81) compared to those who live in other cities. Participants who consider themselves sex workers had four times the odds of having syphilis (OR=2.97, 95\%CI 1.63-9.51) than those whose not consider themselves as sex worker. Participants who had any sexually transmitted infection (STI) symptom in the past year had  three times the odds of having syphilis (OR=3.02, 95\%CI 1.62-5.65) compared to those participants that did not have any symptom of STI in the past year.

\section{Discussion}

In this paper, we present a strategy to model binary response variables from respondent-driven sampling data. Firstly, the network dependence of the response variable should be tested, we propose to test the dependence by building a contingency table with the quantity of interest of the recruiters and the participants and run an independence test. If there is any evidence of network dependence of the response variable, we suggest using the RDS II estimator rather than the naive estimator to estimate the prevalence of the quantity of interest. The binary regression model with latent effects can be an alternative to regression models for RDS data that ignore the network structure.

We observed that there is some evidence on network dependence for HIV and syphilis serostatus. Using the RDS II estimator the prevalences are 7.1\% (4.7; 9.6) for HIV and 9.4\% (1.5; 17.4) for syphilis. 

There are some issues that still need to be addressed. The RDS II estimator relies on the sampling-with-replacement assumption, and the biases introduced due to sampling without-replacement are unknown. Volz and Heckathorn (2008) \cite{VolHec08} discuss this and some other issues related to the use of the RDS II estimator on practice. 
Another important issues not tackled in this paper are the missing data. When there are some missing information the observations with missing were removed and the observed network was broken. Therefore, imputation methods for network data are needed.

The binary regression with the latent network effect assumes that the observed network contains all the information about the social network. However, the network observed from respondent-driven sampling data is incomplete. This is due to the limited number of friends each person can bring and also the fact that the each individual cannot participate more than once in the study. Hence, for future research, we intend to reconstruct the social network of the sample using the degree information and other explanatory variables, and given the estimated social network we could directly apply model (\ref{car}).




\bibliographystyle{vancouver}
\bibliography{../../Papers/RDS.bib} 


\section*{Tables}

\begin{table}[ht]
\caption{\bf{HIV serostatus for the participants and its recruiters.}}
\begin{center} 
\begin{tabular}{rrr}
  \hline
 & \multicolumn{2}{c}{Recruiter} \\
Participant & Negative & Positive \\ 
  \hline
Negative & 478 &  27 \\ 
Positive &  33 &  10 \\ 
   \hline
\end{tabular}
\end{center}
\label{hivdep.tbl}
\end{table}

\begin{table}[ht]
\caption{Syphilis serostatus for the participants and its recruiters.}
\begin{center} 
\begin{tabular}{rrr}
  \hline
 & \multicolumn{2}{c}{Recruiter} \\
Participant & Negative & Positive \\ 
  \hline
Negative & 481 &  49 \\ 
  Positive &  70 &  19 \\ 
   \hline
\end{tabular}
\end{center}
\label{syphdep.tbl}
\end{table}

\begin{table}[ht]
\caption{\bf{Estimated prevalence for HIV and syphilis and the correspondent 95\% confidence interval.}}
\begin{center} 
\begin{tabular}{rcc}
  \hline
 & \multicolumn{2}{c}{Estimator} \\
 & Naive & RDS II \\ 
  \hline
HIV      & 0.0789 (0.0577; 0.1001) & 0.0711 (0.0466; 0.0955) \\
Syphilis & 0.1155 (0.0911; 0.1399) & 0.0944 (0.0146 0.1741) \\
   \hline
\end{tabular}
\end{center}
\label{prev.tbl}
\end{table}

\begin{table}[ht]
\caption{\bf{Estimated effects and 95\% credible intervals for the logistic regression (LogReg) and logistic regression with network structure (NetLogReg) with HIV serostatus as response variable}} 
\begin{center}
\begin{tabular}{rcc}
  \hline
 & LogReg (95\% CI) & NetLogReg (95\% CI) \\ 
  \hline
\textbf{(Intercept)} & -4.2076 (-7.0755, -1.8333) & -5.1342 (-8.4060, -2.3608) \\ 
\multicolumn{3}{l}{\textbf{Educational material in the past two months?} (Yes, No)} \\
No & -0.8593 (-1.7351, -0.0616) & -1.1091 (-2.1311, -0.1953) \\ 
\multicolumn{3}{l}{\textbf{Gender identity} (Male, Transvestite, Others)} \\
Transvestite & 0.2809 (-0.7307, 1.2241) & 0.0062 (-1.3697, 1.2326) \\ 
Others & 0.3668 (-0.7745, 1.3809) & 0.5984 (-0.6983, 1.7596) \\ 
\multicolumn{3}{l}{\textbf{Age category} (<25, $\geq 25$)} \\
$\geq 25$ & 1.2033 (0.4629, 1.9826) & 1.4142 (0.5200, 2.3626) \\ 
\multicolumn{3}{l}{\textbf{Belongs to a gay NGO?} (Yes, No)} \\
No & -0.3553 (-1.3367, 0.7136) & -0.9402 (-2.0846, 0.3129) \\ 
\multicolumn{3}{l}{\textbf{Any physical violence ever against gays and trans?} (yes, No)} \\
Yes & 0.5123 (-0.2135, 1.2313) & 0.7455 (-0.1077, 1.6023) \\ 
\multicolumn{3}{l}{\textbf{Total number of partners in the past 2 months} (0,1, $>$1)} \\
1 & 0.7446 (-1.2697, 3.2865) & 0.8539 (-1.3817, 3.6636) \\ 
$>$1 & 1.3142 (-0.5269, 3.7020) & 1.3213 (-0.7057, 3.9303) \\ 
\multicolumn{3}{l}{\textbf{Consider self as sex worker?} (Yes,No)} \\
No & -0.2374 (-1.2433, 0.8524) & -0.1729 (-1.4574, 1.2522) \\ 
\multicolumn{3}{l}{\textbf{Any symptoms of STI in the past year?} (Yes, No)} \\
No & -0.5148 (-1.2260, 0.2141) & -0.3671 (-1.2284, 0.5270) \\ 
\multicolumn{3}{l}{\textbf{At least college degree?} (Yes, No)} \\
No & 0.7460 (-0.1148, 1.6842) & 1.0666 (0.0346, 2.2212) \\ 
\multicolumn{3}{l}{\textbf{UIAI in the last 2 months?} (Yes, No)} \\
Yes & -0.9568 (-1.8539, -0.1426)& -0.9926 (-2.0101, -0.0802) \\ 
   \hline
DIC & 268.20 & 251.97 \\
$p_D$ & 12.12 & 44.85 \\ \hline
\end{tabular}
\end{center}
\label{hivreg.tbl}
\end{table}

\begin{table}[ht]
\caption{\bf{Estimated effects and 95\% credible intervals for the logistic regression (LogReg) and logistic regression with network structure (NetLogReg) with Syphilis serostatus as response variable.}}
\begin{center}
\begin{tabular}{rcc}
  \hline
 & LogReg (95\% CI) & NetLogReg (95\% CI) \\ 
  \hline
\textbf{(Intercept)} & -1.1234 (-2.6638, 0.3757) & -1.1232 (-2.6634, 0.3762) \\
\multicolumn{3}{l}{\textbf{Educational material in the past two months?} (Yes, No)} \\
No & -0.2420 (-0.9397, 0.4315) & -0.2424 (-0.9402, 0.4311) \\
\multicolumn{3}{l}{\textbf{Gender identity} (Male, Transvestite, Others)} \\
Transvestite & 0.9105 (0.0288, 1.7532) & 0.9104 (0.0287, 1.7532) \\
Others & 0.8312 (-0.0796, 1.6815) & 0.8315 (-0.0790, 1.6822) \\ 
\multicolumn{3}{l}{\textbf{Age category} ($<25$, $\geq 25$)} \\
$\geq 25$ & 1.0733 (0.4334, 1.7449) & 1.0733 (0.4334, 1.7450) \\
\multicolumn{3}{l}{\textbf{Race }(White, Black/mulatto, Other)} \\
Black/mulatto & 0.5918 (-0.0375, 1.2291) & 0.5925 (-0.0367, 1.2300) \\
Other & -0.0630 (-2.5805, 1.9089) & -0.0622 (-2.5790, 1.9121) \\
\multicolumn{3}{l}{\textbf{City of residence} (Campinas, Other)} \\
Other & -0.9087 (-1.9181, -0.0244)  & -0.9099 (-1.9196, -0.0257) \\
\multicolumn{3}{l}{\textbf{Brazilian criterion for purchase power} (A/B/C, D/E )} \\
A/B/C\{richest\} & -0.2900 (-1.1786, 0.6633) & -0.2899 (-1.1785, 0.6635) \\
\multicolumn{3}{l}{\textbf{Belongs to a gay NGO?} (Yes, No)} \\
No & 0.2521 (-0.7466, 1.3456) & 0.2522 (-0.7466, 1.3457) \\
\multicolumn{3}{l}{\textbf{Received free condom?} (Yes, No)} \\
No & -0.8730 (-2.0934, 0.1731) & -0.8728 (-2.0933, 0.1736) \\ 
\multicolumn{3}{l}{\textbf{Consider self as sex worker?} (Yes, No)} \\
No & -1.3785 (-2.2522, -0.4881) & -1.3794 (-2.2534, -0.4891) \\
\multicolumn{3}{l}{\textbf{Any symptoms of STI in the past year?} (Yes, No)} \\
No  & -1.1053 (-1.7312, -0.4846) & -1.1059 (-1.7318, -0.4851) \\ 
\multicolumn{3}{l}{\textbf{Regular drug user?}  (Yes, No)} \\
No  & -0.2307 (-1.0048, 0.5917) & -0.2306 (-1.0047, 0.5919) \\
\multicolumn{3}{l}{\textbf{UIAI in the last 2 months?}  (Yes, No)} \\
Yes & 0.1832 (-0.4584, 0.8197) & 0.1831 (-0.4585, 0.8197) \\
   \hline
DIC & 320.14 & 320.13 \\
$p_D$ & 14.11 & 14.13 \\ \hline
\end{tabular}
\end{center}
\label{syphreg.tbl}
\end{table}


\end{document}